%% file: main.tex
\documentclass[conference]{IEEEtran}
\IEEEoverridecommandlockouts

\usepackage{cite}
\usepackage{amsmath,amssymb,amsfonts}
\usepackage{algorithmic}
\usepackage{graphicx}
\usepackage{textcomp}
\usepackage{xcolor}
\usepackage{url}
\def\BibTeX{{\rm B\kern-.05em{\sc i\kern-.025em b}\kern-.08em
    T\kern-.1667em\lower.7ex\hbox{E}\kern-.125emX}}
\begin{document}

\title{
\footnotesize
\framebox[1.01\width]{\parbox{\dimexpr\linewidth-2\fboxsep-2\fboxrule}{If you cite this paper, please use this reference: B. Thapa et al. \emph{Embedded Rust or C Firmware? Lessons from an Industrial Microcontroller Use Case with Ariel OS}. In Proceeding of IEEE DCOSS-IoT, Workshop on IoT Applications and Industry 5.0 (IoTI5), June 2026.}}
 \ \\ \ \\ \ \\
\Huge
Embedded Rust or C Firmware? Lessons from an \\
Industrial Microcontroller Use Case with Ariel OS
}

\author{
B. Thapa\textsuperscript{1,2}, D. Alfonso\textsuperscript{1}, L. Bini\textsuperscript{1}, L. Mapelli\textsuperscript{1}, K. Schleiser\textsuperscript{3,4}, R. Fouquet\textsuperscript{4}, E. Baccelli\textsuperscript{3,4}\\ \\
\textsuperscript{1} STMicroelectronics, \emph{Italy} \\
\textsuperscript{2} Polytechnic of Turin, \emph{Italy} \\
\textsuperscript{3} Freie Universit\"at Berlin, \emph{Germany} \\
\textsuperscript{4} Inria, \emph{France} 
}

\maketitle

\begin{abstract}
\input{Text/0-abstract}
\end{abstract}


\input{Text/1-introduction}

\input{Text/2-case-study}
\input{Text/3-approach}
\input{Text/4-VDL-C}
\input{Text/5-VDL-Rust}
\input{Text/6-qualitative-comparison}
\input{Text/7-measurements}
\input{Text/8-lessons-learnt}
\input{Text/8-1-related-work}
\input{Text/9-conclusion}

\cite{1} M. Chowdhary et al. "ST AIoT Craft-A No-Code/Low-Code Cloud Solution for Edge AI Management in Smart Sensors." Proceedings of ACM SenSys, 2025.
\cite{2}  E. Frank et al. "Ariel OS: An Embedded Rust Operating System for Networked Sensors \& Multi-Core Microcontrollers." Proceedings of IEEE DCOSS-IoT, 2025.
\cite{3} J.Van Der Stoep, “ Rust in Android: Move Fast and Fix Things ,” in  Google Security Blog, Nov. 2025. Online: \url{https://security.googleblog.com/2025/11/rust-in-android-move-fast-fix-things.html}
\cite{4} ST AIoT Craft. Online: \url{https://staiotcraft.st.com}
\cite{5} STMicroelectronics, “SensorTile.box PRO with multi-sensors and wireless connectivity”, Online: \url{https://www.st.com/resource/en/data_brief/steval-mkboxpro.pdf}
\cite{6} STMicroelectronics, LSM6DSV16X Datasheet DS13510, “6-axis inertial measurement unit”, Online: \url{https://www.st.com/resource/en/datasheet/lsm6dsv16x.pdf}
\cite{7} STM32CubeMX: Stm32 Cube initialization code generator. Online: \url{https://www.st.com/en/development-tools/stm32cubemx.html}
\cite{8} Embassy Framework: \url{https://github.com/embassy-rs/embassy}
\cite{9} STMicroelectronics MEMS Rust Drivers Repository. Online: \url{https://github.com/STMicroelectronics/st-mems-rust-drivers}
\cite{10} DTDL: Digital Twins Definition Language. Online:   \url{https://github.com/Azure/opendigitaltwins-dtdl}
\cite{11} Azure IoT Plug and Play (PnP). Online:  \url{https://learn.microsoft.com/en-us/previous-versions/azure/iot/overview-iot-plug-and-play}
\cite{12} Parson library. Online: \url{https://github.com/kgabis/parson/}
\cite{13} A. Sharma et al. "Rust for embedded systems: Current state and open problems." Proceedings of ACM SIGSAC, 2024.
\cite{14} H. Ayers et al. "Tighten Rust’s belt: shrinking embedded Rust binaries." Proceedings of ACM SIGPLAN/SIGBED LCTES, 2022.
\cite{15} Rust Asynchronous Programming. Online: \url{https://doc.rust-lang.org/book/ch17-00-async-await.html}, 2026.
\cite{16} RTIC. Online: \url{https://github.com/rtic-rs/rtic}, 2026.
\cite{17} A. Levy et al., “Ownership is theft: Experiences building an embedded OS in Rust,” in ACM PLOS, 2015.
\cite{18} Rust Community Crate Registry. Online: \url{https://crates.io}, 2026.
\cite{19} XCube MEMS. Online: \url{https://www.st.com/en/embedded-software/x-cube-mems1.html}, 2026.
\cite{20} Serde Framework. Online: \url{https://serde.rs/}, 2026.
\cite{21} EU Cyber Resilience Act. Online \url{https://www.cyberresilienceact.eu/}
\cite{22} Newlib. Online: \url{https://sourceware.org/newlib/}
\cite{23} \noindent Heapless crate. Online: \url{https://github.com/rust-embedded/heapless}
\cite{24} X-NUCLEO-IKS4A1 MEMS expansion board for STM32 Nucleo. Oonline: \url{https://www.st.com/en/evaluation-tools/x-nucleo-iks4a1.html}
\cite{25} Bloaty. Online: \url{https://github.com/google/bloaty}
\cite{26} Cargo-bloat. Online: \url{https://github.com/RazrFalcon/cargo-bloat}
\cite{27} StaticCell. Online: \url{https://github.com/embassy-rs/static-cell}
\cite{28} 

\end{document}

%% file: Text/0-abstract.tex
As Rust gains traction for developing safer systems software, a reality check for the microcontroller hardware segment becomes necessary. How ready is the Rust ecosystem for this segment? Can Rust compete with C in practice? This paper reports on an IoT industrial case study that contributes to answering these questions. We analyze two teams concurrently developing the same functionality — one in C, one in Rust — over a period of several months. A comparative analysis of their approaches, results, and iterative efforts is provided. The analysis and measurements on hardware indicate no strong reason to prefer C over Rust for microcontroller firmware on the basis of memory footprint or execution speed. Furthermore, Ariel OS is shown to provide an efficient and portable system runtime in Rust whose footprint is smaller than that of the state-of-the-art bare-metal C stack traditionally used in this context. It is concluded that Rust is a sound choice today for firmware development in this domain.

%% file: Text/1-introduction.tex
\section{Introduction}
A trend is emerging in embedded software for networked microcontrollers, where Rust is increasingly championed as a safer alternative to C/C++, which has predominated in firmware development for this segment. While a large-scale study on Android firmware has documented the feasibility and benefits of transitioning from C/C++ to Rust [3] on less resource-constrained hardware, the state of the art for embedded Rust on microcontrollers is comparatively less established. The overarching question addressed in this paper is therefore: how ready is embedded Rust in practice as a replacement for traditional C/C++ implementations?

With this purpose in mind, a case study analyzing an IoT industrial use case is reported. A large vendor's product line spanning heterogeneous microcontroller hardware used as sensor data loggers is considered. Parallel firmware development efforts in C and Rust — each providing the required functionality — are analyzed and compared.

During {\bf Phase 1}, both teams worked in isolation for $\approx$6~weeks, at the end of which each delivered a functional firmware — VDL-C and VDL-Rust respectively. During  {\bf Phase 2}, over an additional $\approx$4~weeks, the two teams met on a weekly basis to review performance and discuss incremental improvements to both implementations. This paper reports on these results and on the iterative efforts made during Phase 2.
More in details, this paper describes the following contributions:
\begin{itemize}
    \item We provide a case study of the STAIoTCraft data logger, an edge AI toolkit targeting heterogeneous microcontrollers and sensors;
    \item We report on parallel efforts designing, implementing, and optimizing VDL-C and VDL-Rust — two firmware implementations of the same STAIoTCraft functionality, in C and Rust respectively;
    \item We provide a comparative analysis of the architectures of VDL-C and VDL-Rust;
   \item We provide comparative performance measurements of VDL-C and VDL-Rust running on commercial MCU hardware;
   \item We conclude with an assessment of the readiness of Rust for microcontroller firmware development.
\end{itemize}

%% file: Text/2-case-study.tex
\section{Case Study: STAIoTCraft}
\begin{figure}[t]
\centering
\includegraphics[width=\columnwidth]{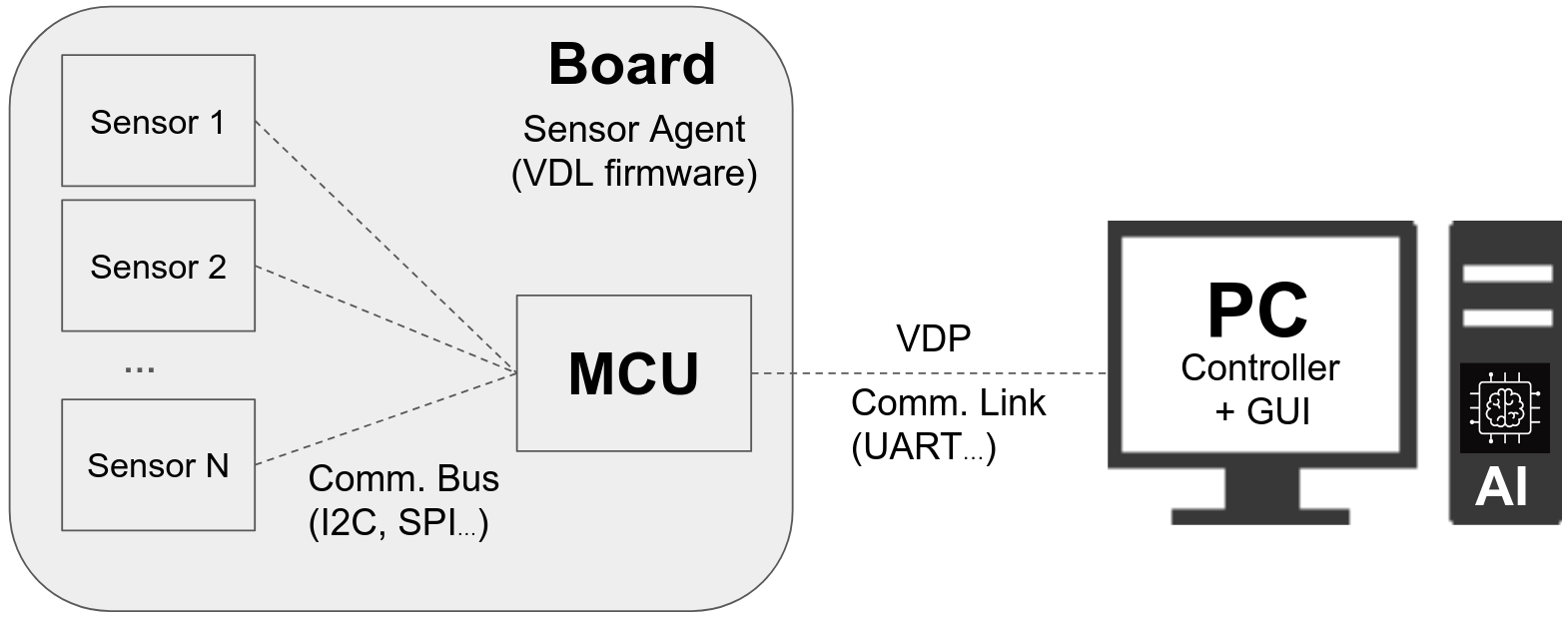}
\caption{STAIoTCraft: VDL firmware on an MCU interacting with a PC controller via VDP protocol over a communication link, e.g., UART.}
\label{fig:STIoTCraft-Diag}
\end{figure}

STAIoTCraft is an online platform and a toolkit provided by STMicroelectronics for training, deploying, and managing distributed artificial intelligence of things (AIoT) devices, gateways, and cloud infrastructure from a web browser, described in prior work [1], and available online [4]. The system has two main components shown in Fig.~\ref{fig:STIoTCraft-Diag}.  On the one hand, a {\bf sensor agent} captures sensor data in real time, and streams this data via a communication link. Correspondingly, a {\bf controller} receives sensor data streams, trains a model and/or displays sensor data and in-sensor inference results in a browser GUI (running on a PC);

Sensor data is uploaded to the cloud to build a classified dataset used to train a machine learning model. In a second phase, the model is then fed back to the sensor agent to perform in-sensor inference and classification of data.

{\bf Sensor Agent Hardware –} The reference hardware used as sensor data logger with the STAIoTCraft toolkit is the SensorTile.box Pro [5], integrating a STM32U585AI microcontroller unit (MCU) based on Arm Cortex-M33 core and a LSM6DSV16X 6-axis MEMS inertial measurement unit (IMU) with embedded machine learning core. However, the toolkit allows the use of a wide variety of devices, microcontrollers, environmental and motion MEMS sensors, Time-of-Flight sensors… 

{\bf Sensor Agent Software –} The firmware running on the sensor agent is thereafter referred to as the Vanilla Data Logger (VDL) firmware. The protocol used by the sensor agent to communicate sensor data to the controller is the Vanilla Datalog Protocol (VDP), detailed below. 

{\bf Vanilla Datalog Protocol (VDP) –} This protocol is used for sensor data logging and serves as the de facto common denominator to support heterogeneous sensor agent hardware. VDP enables interaction between a logging controller and a sensor agent (recall Fig. 1).  VDP is used to transmit data over a communication link, such as a UART peripheral, and it is based on three layers:

\begin{itemize}
    \item {\bf Asymmetric Serial Packet Exchange Protocol (ASPEP)}: Layer specifying semantics and syntax of packets headers;
    \item {\bf Simple Serial Transport Layer (SSTL)}: This layer specifies a point-to-point, master/slave protocol that allows data transfer on a serial link, e.g. a UART peripheral. Both synchronous and asynchronous transmissions are supported. SSTL is tailored for multi-channel data streaming;
    \item {\bf Data Layer}: This layer carries command-response exchanges and the sensor data to be transmitted.
\end{itemize}

The command-response protocol and binary data format are described in device models generated using DTDLv2 [10], a JSON-based language for describing digital twins. These models are used within the Azure IoT Plug and Play (PnP) framework [11], which STAIoTCraft adopts for datalogging. The firmware is responsible for handling JSON message de/serialization and streaming sensor data at the configured Output Data Rate (ODR).

%% file: Text/3-approach.tex
\section{Approach \& Methodology}

In the following, we focus on a comparative evaluation of two firmware implementations, one in C and one in Rust, that each provide convenient hardware abstraction, comply with STAIoTCraft requirements, and notably implement VDP. We use the following terminology: {\bf VDL-C is the firmware developed in C, while VDL-Rust is the Rust firmware}.  

VDL-C and VDL-Rust were developed by separate teams, within a comparable timeframe ($\approx$10 weeks). Each team was tasked to provide a VDP-compliant firmware able to stream sensor data from a sensor agent to the STAIoTCraft controller as described in the previous section. Both teams were missioned with a common out-of-the-box mindset: aim to provide a simple, easy-to-understand, easy-to-use, and easy-to-customize STAIoTCraft data logger firmware example.  In a nutshell, the focus is on one-size-fits all: portability and maintainability of the code.

While sections IV, V and VII focus on concrete sensors and a specific microcontroller to satisfy STAIoTCraft requirements, sections VI, VIII and IX aim to discuss more general observations that are not limited to this scope, applicable to more diverse microcontrollers and MEMS.

%% file: Text/4-VDL-C.tex
\section{C Firmware Overview: VDL-C}

\begin{figure}[t]
\centering
\includegraphics[width=\columnwidth]{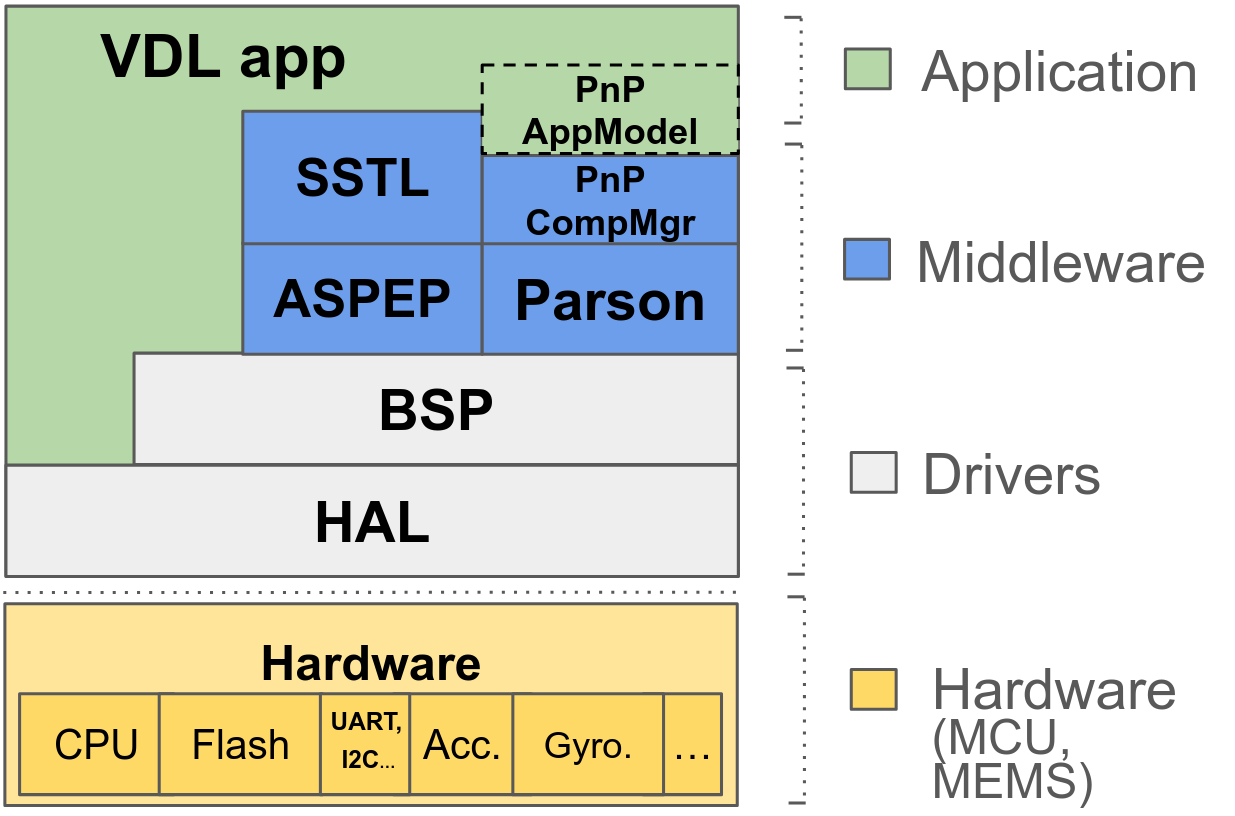}
\caption{Architecture of the VDL-C firmware stack.}
\label{fig:VDL-C-arch}
\end{figure}

To facilitate our C implementation of the VDP protocol and its portability across heterogeneous microcontroller hardware, we extensively use state-of-the-art tooling for code autogeneration and integrate open source building blocks. The VDL-C firmware stack is displayed in Fig. \ref{fig:VDL-C-arch}.

{\bf Use of CubeMX —} The board support package (BSP), hardware abstraction layer (HAL), and hardware initialization code are auto-generated using STM32CubeMX [7], a de facto standard tool with a convenient graphical user interface (GUI). As a first step, STM32CubeMX is used to generate an initial project containing all hardware initialization files in C, from which the target board can be selected (e.g., ST SensorTile.BoxPro), the needed peripherals configured, and the desired software packages added — such as the CMSIS X-CUBE-MEMS1 function pack for sensor support and drivers [19]. As a second step, the C code template for PnP components and the corresponding glue code for the middleware shown in Fig. \ref{fig:VDL-C-arch} are generated.

\begin{figure}[t]
\centering
\includegraphics[width=\columnwidth]{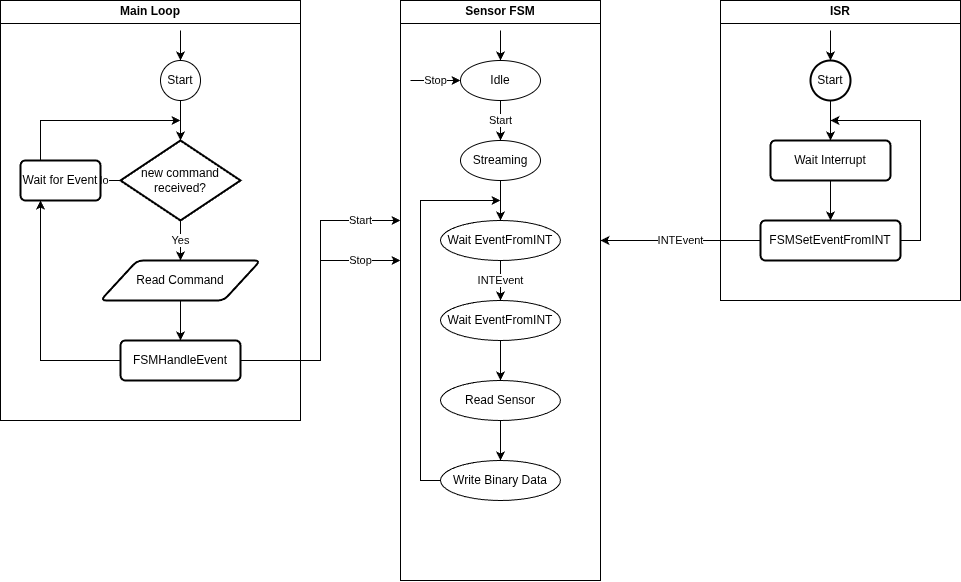}
\caption{VDL-C firmware stack tasks and IPC.}
\label{fig:VDL-C-IPC}
\end{figure}

{\bf Finite State Machine \& Interrupt Events —} The VDL-C firmware is written in bare-metal C as an event-driven application. Its architecture follows a classical Finite State Machine (FSM) running within a main endless loop, as depicted in Fig. \ref{fig:VDL-C-IPC}. The FSM handles events from the LSM6DSV16X sensor interrupts, the UART/I²C communication backend, and the MCU timers. Built on top of the standard CubeMX-generated code, VDL-C implements FSM reactions to commands received via the VDP protocol, handles interrupt events generated by sensors and timers, and exposes VDP protocol functions over the UART serial port.

{\bf Serializing/Deserializing JSON —} VDL-C bundles the open-source Parson library [12] for JSON de/serialization. This is the only firmware component that requires dynamic heap memory allocation (malloc and free), supplied by newlib [22], the C standard library included with the ARM GCC toolchain.

%% file: Text/5-VDL-Rust.tex
\section{Rust Firmware Overview: VDL-Rust}

To facilitate the implementation of the VDP protocol and its portability across heterogeneous microcontroller hardware, VDL-Rust extensively relies on established building blocks from the Rust crate ecosystem, as described below. Building on this ecosystem, VDL-Rust is designed as a business logic layer leveraging APIs provided by Ariel OS [2], a lightweight open-source library operating system for microcontrollers written in Rust.

\begin{figure}[t]
\centering
\includegraphics[width=0.88\columnwidth]{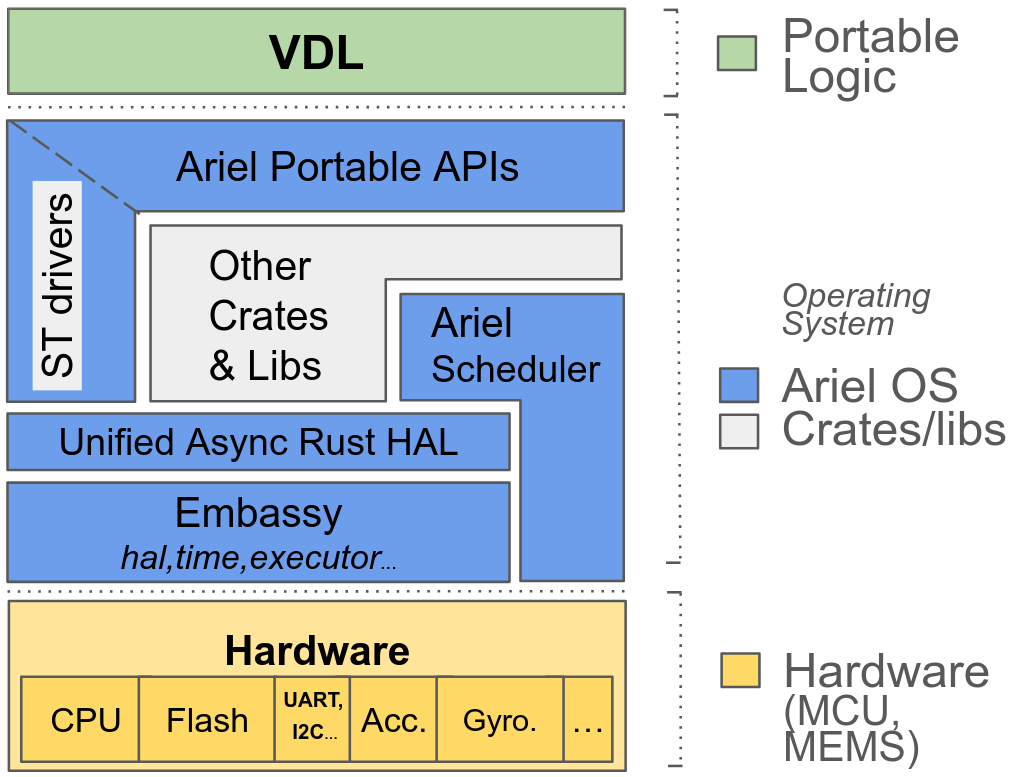}
\caption{Architecture of the VDL-Rust firmware stack.}
\label{fig:VDL-Rust-arch}
\end{figure}

{\bf Use of Ariel OS —} Ariel OS provides a hardware abstraction layer (HAL) built on Embassy [8], suitable for Rust implementations under no\_std and no\_alloc constraints — meaning no underlying OS is assumed and only static memory is allocated. The modular designs of both Ariel OS and the embedded Rust crate ecosystem enable fine-grained feature selection at build time, minimizing resource usage. Ariel OS thus provides portable abstraction and initialization for diverse MCUs, sensors, and communication buses (UART, I²C, SPI, etc.). Within this framework, VDL-Rust imports the required ST sensor drivers [9] from crates.io. An overview of this architecture is shown in Fig.~\ref{fig:VDL-Rust-arch}; for further details on Ariel OS, the reader is referred to [2].

{\bf Serialize/Deserialize with Static Memory —} A pivotal part of VDP logic is JSON command-response handling. VDL-Rust combines the serde crate [20], which provides a serialization/deserialization framework, with the heapless crate [23], which provides types that mirror dynamic memory equivalents — such as Vec and String — but operate on fixed-capacity static memory. To meet the static memory allocation requirement, the serde-json-core crate [28] is used. A nested struct matching the device configuration returned in response to the get\_status command is implemented, along with a nested enum deserialized from incoming JSON commands. After deserialization, Rust's pattern matching makes it straightforward to dispatch on command variants implementing VDP.

\begin{figure}
\centering
\includegraphics[width=0.9\columnwidth]{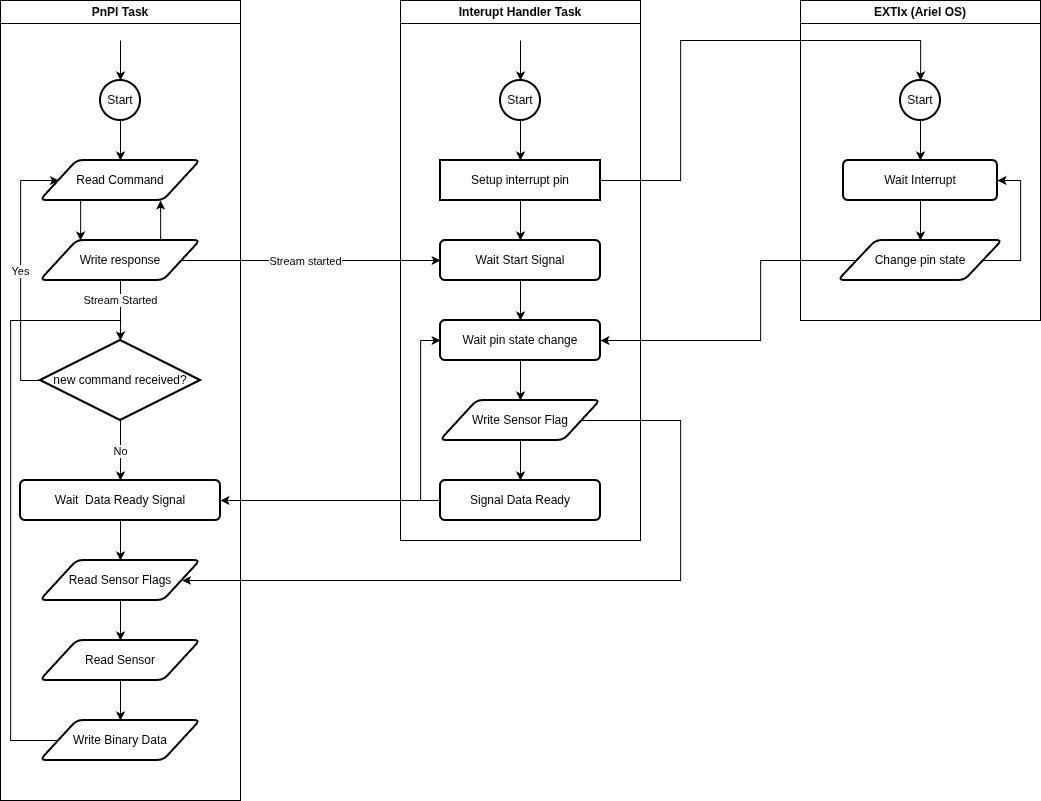}
\caption{VDL-Rust firmware stack tasks and IPC.}
\label{fig:VDL-Rust-IPC}
\end{figure}

{\bf Sensor Data Streaming \& IPC —} Sensor reads are initiated by data-ready interrupts. Tasks corresponding to each interrupt monitor pin state changes on the external interrupt channels. For inter-process communication (IPC), any interrupt-waiting task can send a data-ready signal to the streaming task. Upon receiving a signal, the streaming task checks each flag to identify which sensors have triggered and reads the corresponding data. While streaming is active, the streaming task also checks whether a stop command has been received from the controller application. IPC and tasks are depicted in Fig.~\ref{fig:VDL-Rust-IPC}. VDL-Rust leverages the embassy-sync crate for inter-task communication, which provides thread-safe primitives such as signals and channels. For simple boolean flags, atomic types are used instead, as they guarantee atomic access and satisfy the compiler's memory safety requirements.

%% file: Text/6-qualitative-comparison.tex
\section{Comparing Firmware Design}

Both VDL-C and VDL-Rust adopt a typical event-driven approach, where sensor hardware interrupts trigger I²C reads. Although a polling-based alternative exists, a discussion of that tradeoff falls outside the scope of this paper. Instead, we focus on key differences observed in the design of VDL-Rust compared to VDL-C.

{\bf Use of an RTOS —} The VDL-C architecture is generated by STM32CubeMX, which provides library functionality to configure the MCU and access its peripherals via high-level C APIs. The resulting firmware includes auto-generated, MCU-specific initialization code, after which the application entry function is called. While high-level C APIs for peripheral access are provided, no threading, IPC, or synchronization primitives are available. Concurrency is managed by manually enabling and disabling IRQs.

In contrast, VDL-Rust application code includes no MCU-specific initialization or setup code, relying instead on Ariel OS, which brings up the MCU, spawns the application tasks, and provides a complete set of IPC and synchronization primitives — enforced at the type level by the Rust compiler.

Specifying MCU setup as done in VDL-C provides maximum control but hinders portability. VDL-C can be ported by creating a new STM32CubeIDE project and copying over the application logic, which uses conditional compilation to support multiple MCU types. In principle, the application logic could be factored out into a reusable module; however, maintainability would typically become brittle as the number of supported hardware targets grows, due to the proliferation of distinct STM32CubeIDE projects required.

In contrast, VDL-Rust benefits from Ariel OS escape hatches to override the default MCU configuration where necessary, and porting to another board is straightforward thanks to Ariel OS' portable build system. This level of portability, however, may come at a cost in terms of memory footprint and/or execution speed. Comparative measurements to evaluate this tradeoff are provided in the following section.

{\bf Use of dynamic memory —} While VDL-Rust relies exclusively on static memory allocation, VDL-C uses dynamic heap memory (malloc and free) in its JSON parser (Parson, used by the PNPL code generator by default). As memory usage minimization was not a key goal, we kept the allocating JSON parser, noting that there are less memory heavy options that do not require a heap at all. 
Corresponding memory footprint measurements are provided in the following sections.

{\bf State handling —} An event-driven architecture requires managing state across multiple interrupt and input sources. While VDL-C implements a hand-crafted state machine, VDL-Rust leverages the async/await functionality [15] offered by Rust. Although Rust provides well-documented advantages in terms of code simplicity and correctness [3], the binaries produced by the Rust compiler with async/await warrants performance verification. In the following section, an execution speed penalty detected via logic analyzer is discussed, attributed to the use of async/await in Rust.

{\bf Use of crates —}  VDL-C relies on the STM32CubeMX native packaging mechanism for code reuse, whereas VDL-Rust uses the Rust crate ecosystem. While the former enables reuse within the STM32Cube ecosystem — limited to STM32 MCU applications — the latter extends reuse across application domains, including embedded, WebAssembly, and desktop. For instance, the serde crate [20] used for JSON handling is the de facto standard in the broader Rust ecosystem, and is therefore mature and battle-tested. Furthermore, VDL-Rust factors out the PnP protocol code into a separate crate, facilitating host-side testing independent of the embedded stack, and enabling reuse in WebAssembly or desktop applications. This same crate was reused in the benchmark application used to produce the measurements presented in this paper, covering both VDL-C and VDL-Rust.

%% file: Text/7-measurements.tex
\section{Measurements on Hardware}

In this section, performance measurements of VDL-C and VDL-Rust binaries running on hardware are reported. For brevity and comparability, measurements are presented for the SensorTile.box Pro [5] only, though both firmwares also support various MEMS hardware configurations. The SensorTile.box Pro is based on an ARM Cortex-M33 microcontroller (STM32U585AII6Q) clocked at 160 MHz, with 2 MiB of Flash (ROM) and 786 KiB of SRAM. The measurements cover a VDP implementation targeting the LSM6DSV16X sensor only, using its accelerometer, gyroscope, and machine learning core (MLC) components.

\subsection{Memory Footprint}

\begin{figure}[t]
\centering
\includegraphics[width=\columnwidth]{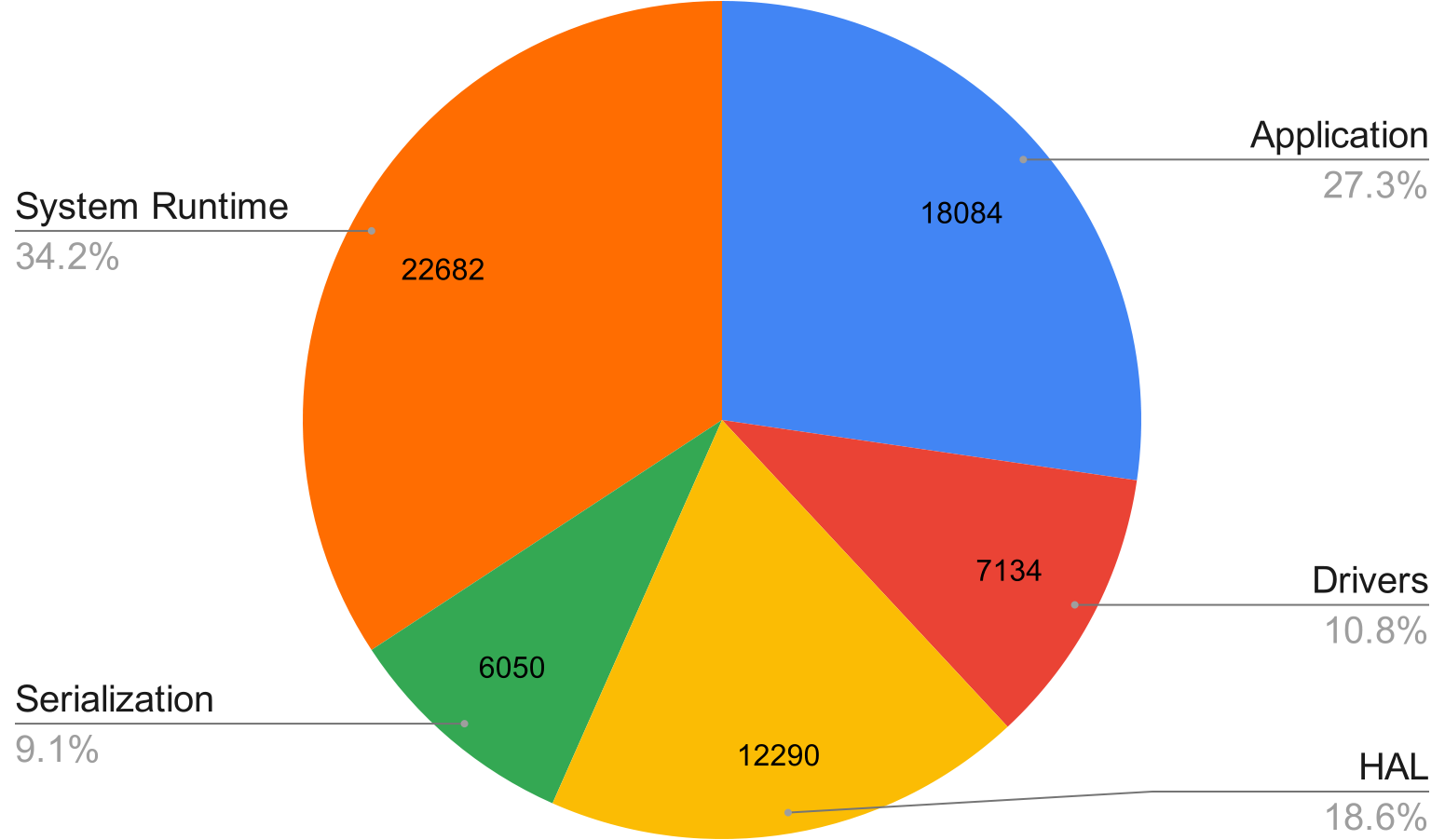}
\caption{VDL-C firmware binary breakdown on Cortex-M33.}
\label{fig:VDL-C-bin-breakdown}
\end{figure}

\begin{figure}[t]
\centering
\includegraphics[width=\columnwidth]{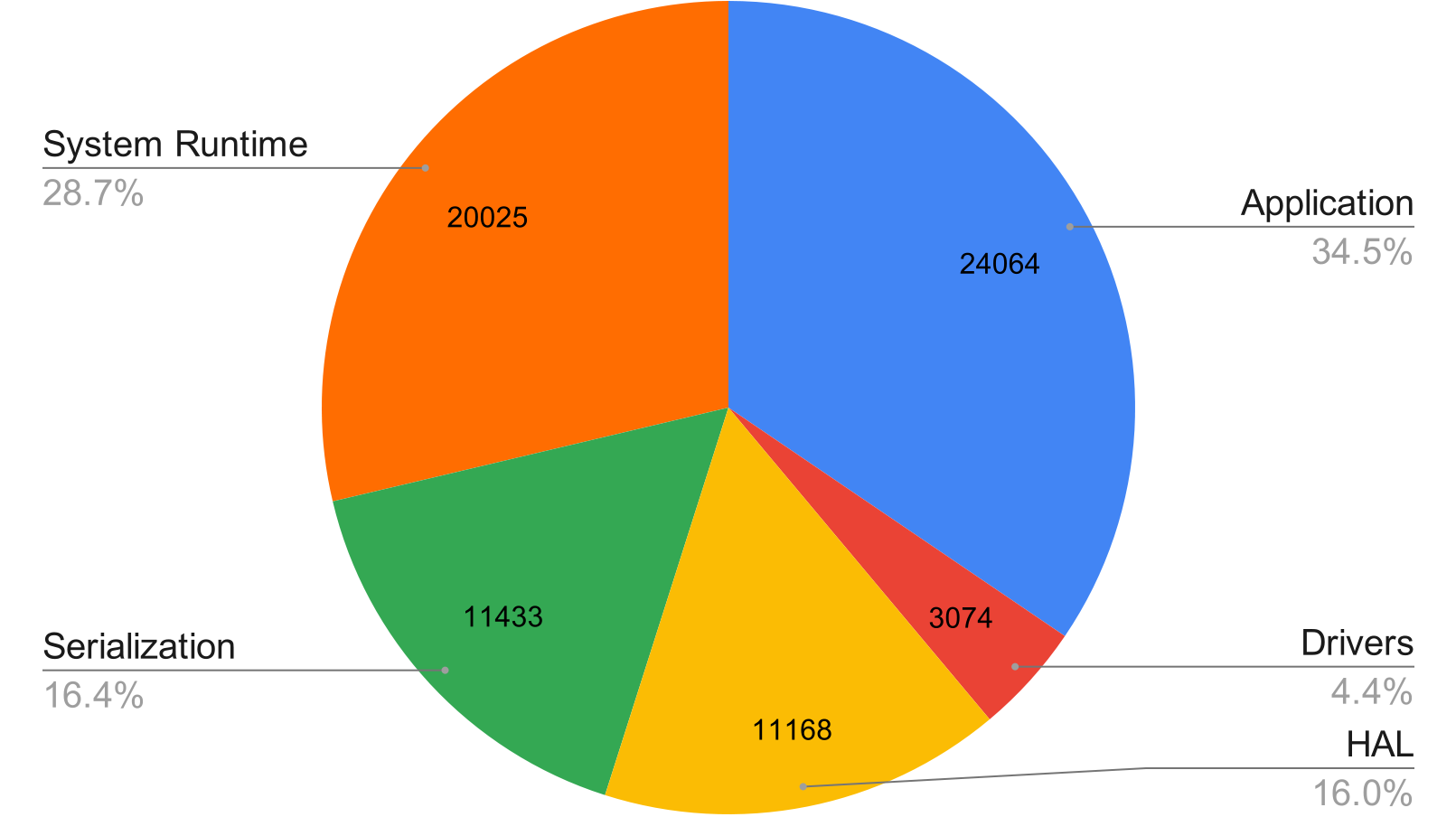}
\caption{VDL-Rust firmware binary breakdown on Cortex-M33.}
\label{fig:VDL-Rust-bin-breakdown}
\end{figure}

A breakdown of the VDL-C and VDL-Rust firmware is shown in Fig.~\ref{fig:VDL-C-bin-breakdown} and Fig.~\ref{fig:VDL-Rust-bin-breakdown}, respectively. Both implementations were profiled for code size and RAM usage; results are summarized in Table~\ref{tab1}.

{\bf Total Code Size} (.text section). Measured using bloaty [25] and cargo-bloat [26], the .text section totals 66,240 bytes for VDL-C and 69,764 bytes for VDL-Rust (~5\% larger). Of VDL-C's total, 21,389 bytes (32.3\%) are occupied by newlib, the toolchain-bundled C standard library on which Parson relies for malloc and free.  Total ROM (including .rodata) stands at 84,100 bytes for VDL-Rust and 76,744 bytes for VDL-C (~10\%), consistent with the .text comparison.

{\bf Application Code Size.} Excluding system runtime code size (only looking at Application, Drivers and Serialization in Fig.~\ref{fig:VDL-C-bin-breakdown} and Fig.~\ref{fig:VDL-Rust-bin-breakdown}), VDL-Rust is larger by ~23\%, driven by application logic (24,064 vs. 18,084 bytes) and serialization (11,433 vs. 6,050 bytes), the latter a consequence of serde's compile-time monomorphization.

{\bf System Runtime Code Size.}  Looking at system runtime size (System + HAL in Fig.~\ref{fig:VDL-C-bin-breakdown} and Fig.~\ref{fig:VDL-Rust-bin-breakdown}),  VDL-Rust shows ~10\% smaller code. For VDL-C, this mainly accounts for newlib + startup code. For VDL-Rust this accounts for Ariel OS + startup code. This observation is surprising because one would expect that an approach based on an operating system (as VDL-Rust is based on Ariel OS) would yield a heftier price in terms of footprint, compared to an approach closer to bare-metal (such as VDL-C).

{\bf RAM and Stack Usage.} In Table~\ref{tab1}, we show comparative RAM measurements. VDL-C uses a 2,048 B stack, 14,960 B of static memory, and a measured peak of 25,600 B of heap, totalling 44,656 B. VDL-Rust uses a 10,240 B stack, 14,400 B of static memory, and no heap, totalling 24,640 B (~45\% less). The heap is entirely attributable to Parson's dynamic allocation of JSON tree nodes; serde-json-core eliminates dynamic allocation by deserializing directly into typed stack structs, so the difference reflects ecosystem conventions as much as deliberate optimization. The larger VDL-Rust stack is not inherent to the language but stems from the async executor model, discussed in Section VIII. VDL-C's total is a measured runtime peak contingent on test coverage; VDL-Rust's is a static compile-time bound. Overall, from a memory footprint point of view, VDL-Rust compares favourably to VDL-C.

\begin{table}[t]
\begin{center}
\caption{Stack \& heap with VDL-C and VDL-Rust on Cortex-M33.}
\label{tab1}
\begin{tabular}{| c | c | c | c |}
\hline
{\bf Metric (bytes)} & {\bf VDL-C} & {\bf VDL-Rust} & {\bf $\Delta$(Rust - C)}\\
\hline
\hline
.text & 66,240 & 69,764 &  3,524\\
\hline
{\bf Total ROM} & {\bf 76,744} & {\bf 84,100} &  {\bf 7,356}\\
\hline
\hline
Stack RAM & 2,048 & 10,240 &  8,192\\
\hline
Static RAM & 14,960 & 14,400 &  -560\\
\hline
Heap RAM & 25,600 & 0 &  -25,600\\
\hline
{\bf Total RAM}  & {\bf 42,608} & {\bf 24,640} &  {\bf -17,968}\\
\hline
\end{tabular}
\end{center}
\end{table}

\subsection{Towards Smaller Memory Footprints}
\begin{figure}[t]
\centering
\includegraphics[width=\columnwidth]{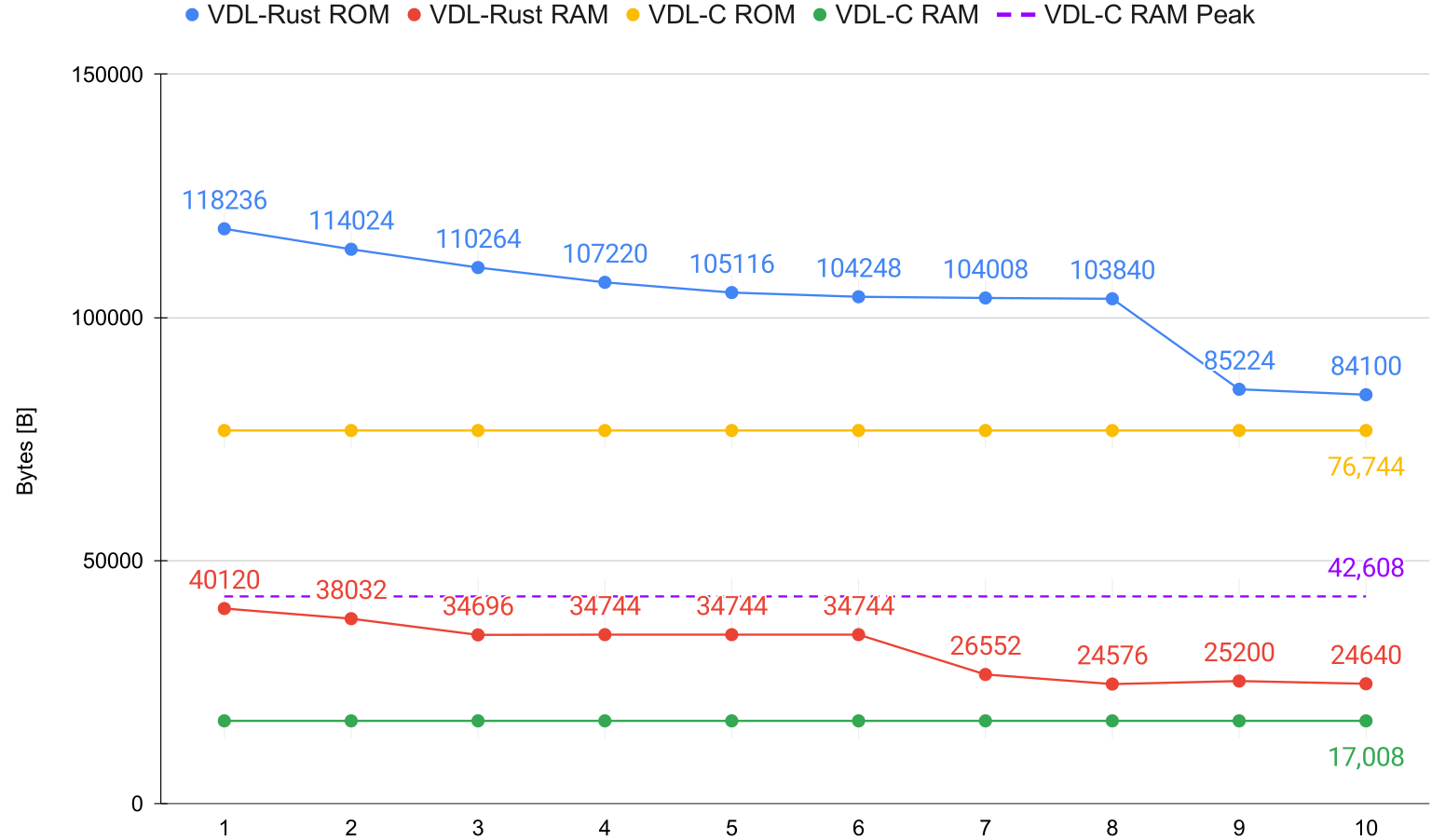}
\caption{Iterations of VDL-Rust RAM/ROM footprint, versus VDL-C.}
\label{fig:mem-footprint-evolution}
\end{figure}

While the previous measurements show the final results, we show in Fig.~\ref{fig:mem-footprint-evolution} an evolution of the footprints in terms of RAM and Flash over the several iterations (during Phase 2). Note that several improvement steps were also necessary for VDL-C, but for brevity, we do not discuss those in this section and in Fig.~\ref{fig:mem-footprint-evolution} we only show the final VDL-C measurements as reference.

 Fig.~\ref{fig:mem-footprint-evolution} depicts ten incremental optimization steps that were applied to reduce the ROM and RAM footprint of VDL-Rust, benchmarked against VDL-C compiled with LTO and -Os (ROM: 76,744 B, RAM: 17,008 B). All subsequent comparisons are made against this baseline.

 By step 10, ROM had been reduced from 118,236 B to 84,100 B — within 10\% of the VDL-C reference (76,744 B). Going in more details, the dominant sources of bloat in VDL-Rust were identified as monomorphization overhead and async state machine inflation. 

{\bf Impact of Monomorphization –} Overhead due to monomorphisation was addressed in steps 3, 4, and 6. These steps consisted in minimizing distinct heapless string types, consolidating per-category enums, and preferring \&str matching over derived serde implementations on enum variants. Each step reduced binary size without altering logic.

{\bf Size of Async Rust State Machine –} Reducing the size of the state machine was addressed in steps 2, 5, 7, 9, and 10.  These steps consisted in passing large buffers by reference (2) and deserializing into compact intermediate enums before the final await point (5) reduced the data retained in machine state. Once large variables were no longer held across await points, the previously necessary 8,192-byte stack increase was exactly reverted (7), recovering 8,192 B of RAM. Replacing async I²C reads with their blocking equivalents (9) eliminated multiple await points across the application and driver layers, and subsequently allowed the PNPL functions to be made fully synchronous (10).

{\bf Overallocation –} The single largest ROM reduction occurred at step 8, where worst-case fixed array sizes were replaced with compile-time configurable sizes, analogous to C preprocessor defines, eliminating approximately 18,000 B through avoided over-allocation.

\begin{figure}[t]
\centering
\includegraphics[width=\columnwidth]{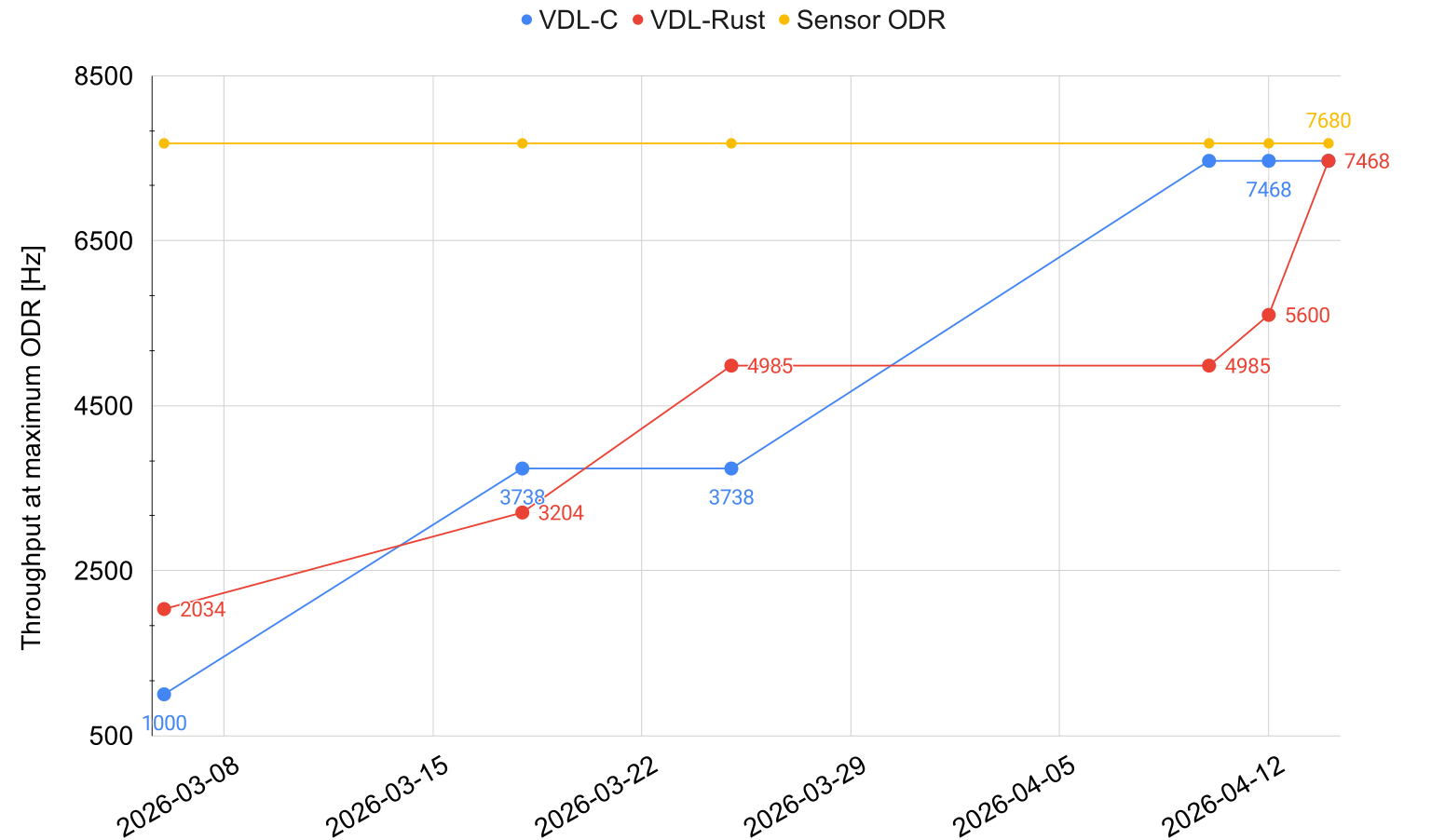}
\caption{Iterations of max. achieved throughput (in Hz) vs target ODR.}
\label{fig:mem-speed-evolution}
\end{figure}

\subsection{Execution Speed}

To gauge execution speed, we measure and compare the time to acquire and transmit sensor data at a selected Output Data Rate (ODR), which is the hardware configuration parameter that defines the rate at which a sensor samples and outputs the data. Whenever a sample is ready, the sensor raises a hardware interrupt, which can be processed by the MCU. For these experiments we used the LSM6DSV16X accelerometer, which has a selectable ODR range from 15 Hz to 7680 Hz. 

\begin{figure*}[t]
\centering
\includegraphics[width=0.7\textwidth]{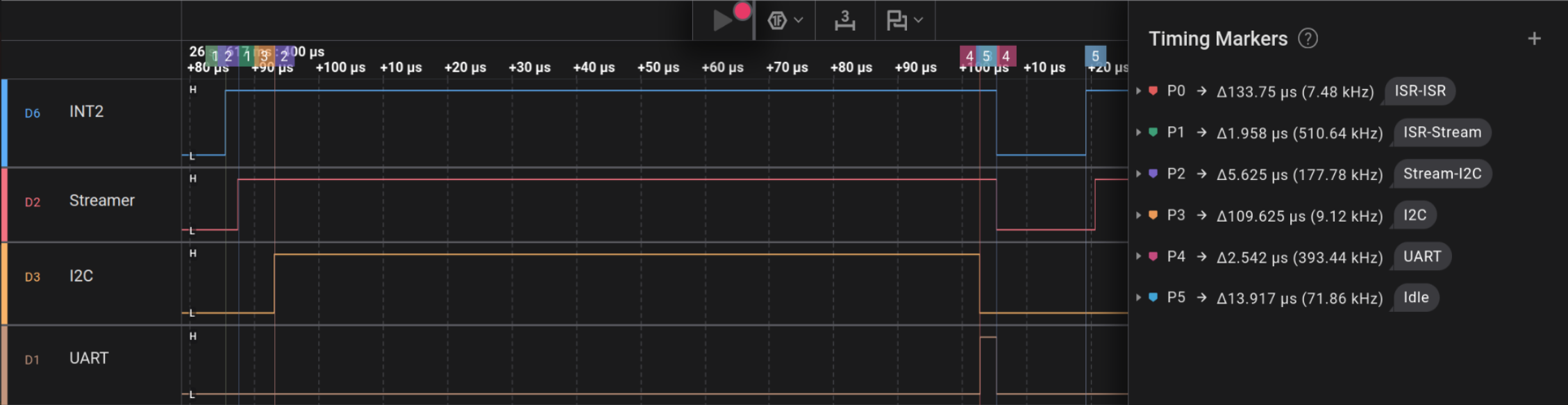}
\caption{Logic analyzer trace of VDL-C with I²C.}
\label{fig:la-C}
\end{figure*}

\begin{figure*}[t]
\centering
\includegraphics[width=0.7\textwidth]{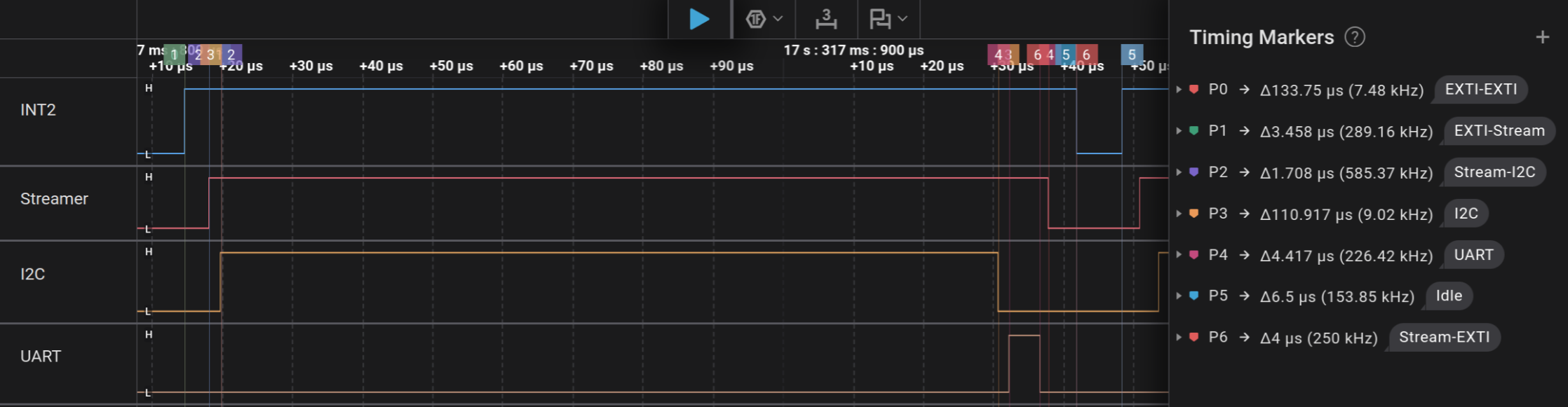}
\caption{Logic analyzer trace of VDL-Rust with I²C.}
\label{fig:la-Rust}
\end{figure*}

Both VDL-C and VDL-Rust firmware successfully achieved servicing the sensor interrupt configured at the maximum ODR of 7680 Hz. As shown in Fig.~\ref{fig:la-C} and Fig.~\ref{fig:la-Rust}, both implementations achieve an interrupt period of 133.75 µs which corresponds to 7477 Hz (see Fig.~\ref{fig:la-C}: P0, Fig~\ref{fig:la-Rust}: P0). The negligible deviation from the target frequency (7680 Hz) can be attributed to sensor-side interrupt drift rather than firmware limitations.

{\bf Impact of async Rust –} A key distinction here between VDL-C and VDL-Rust is the async executor overhead inherent to the async Rust runtime. In VDL-Rust, the hardware interrupt pin is monitored by an Embassy ISR which, upon triggering, wakes the EXTI handler task, which then signals the streaming task. This two-stage dispatch introduces a scheduling latency at the entry point of every cycle that propagates across all subsequent task transitions. This is evidenced by the increased EXTI-to-Stream latency (see Fig.~\ref{fig:la-C}: P1 = 1.958 µs compared to Fig.~\ref{fig:la-Rust}: P1 = 3.458 µs) and UART transition latency (see Fig.~\ref{fig:la-C}: P4 = 2.542 µs compared to Fig.~\ref{fig:la-Rust}: P4 = 4.417 µs), as well as an additional Stream-EXTI transition exclusive to VDL-Rust (see Fig.~\ref{fig:la-Rust}: P6 = 4 µs), representing the executor round-trip back to the EXTI handler awaiting the next interrupt. The cumulative effect of this distributed overhead is reflected in the reduced idle window (see Fig.~\ref{fig:la-C}: P5 = 13.875 µs compared to Fig.~\ref{fig:la-Rust}: P5 = 6.5 µs). Despite this overhead, both implementations meet the interrupt servicing deadline, confirming that though the Rust async execution model incurs some overhead, it does not compromise real-time sensor data acquisition at maximum ODR, albeit at the cost of reduced scheduling margin.

\subsection{Towards Faster Execution}

 Section VII.D. measurements result from iterative improvements that both the VDL-C and the VDL-Rust team performed during Phase 2, that we show in Fig.~\ref{fig:mem-speed-evolution}. 

 Early expectations held that VDL-C would outperform VDL-Rust in terms of execution speed, yet initial measurements yielded the opposite, as shown in Fig.~\ref{fig:mem-speed-evolution}: VDL-Rust peaked at 2034 Hz while VDL-C peaked at 1000 Hz. The disparity was architectural — VDL-Rust initially used polling with non-blocking UART while VDL-C was interrupt-driven with blocking UART writes. Switching VDL-C to non-blocking immediately corrected this.

VDL-Rust's numbers, subsequently rearchitectured to also run interrupt-driven, remained below expectations until logging, which is enabled by default in Ariel OS, was disabled, as it was unneeded for that application. Its removal, alongside several minor fixes, restored the true baseline and briefly put VDL-Rust back ahead at 4985 Hz while VDL-C plateaued at 3738 Hz. VDL-C regained the lead by adopting the same fix VDL-Rust already used — resolving interrupt pin ambiguity between the LSM6DSV16X accelerometer and gyroscope via saved configuration rather than a runtime DRDY status read, eliminating the extra I²C transaction — and from that point sustained the maximum ODRs reliably.

With VDL-C stable, attention turned to the remaining gap in VDL-Rust. Even on a single I²C query, transaction latency continued to be the bottleneck, and closing it required porting each tuning step directly from VDL-C's already-configured values. STM32CubeMX-computed TIMINGR values tightening I²C timing to minimum bus-compliant margins, bringing transaction latency from 122 µs to 112 µs. Enabling I-Cache and flash prefetch then eliminated instruction fetch stalls and pipeline wait states, cutting UART latency from 17 µs to 11 µs and narrowing the stream interval from 138 µs to 126 µs. Disabling the UART FIFO reduced UART latency further still, to 5 µs. Collectively, these steps reduced the streamer task duration from ~140 µs to ~120 µs — an ~20 µs end-to-end improvement.

By week 4 (end of Phase 2) both VDL-C and VDL-Rust reached the same throughput ceiling; the residual discrepancy between the observed and nominal ODR is attributable to sensor-side ODR timing inaccuracies rather than to firmware.

%% file: Text/8-lessons-learnt.tex
\section{Lessons Learnt \& Perspectives}

While C firmware development has been a well-established practice for decades, Rust firmware development for microcontrollers is comparatively recent. Nevertheless, both the C and Rust teams completed their respective tasks within the same short timeframe as mandated in Phase 1. Both teams also required additional effort to improve performance in Phase 2. For full fairness, these observations must be put into perspective: the Rust team was substantially less experienced than the C team, and while the latter leveraged a mature vendor stack, the former relied exclusively on relatively new open-source building blocks and workflows.

The following summarizes general observations distilled from this study, with particular focus on avoidable pitfalls to which developers new to embedded Rust are prone.

\subsection{Code Portability Aspects}

Embedded Rust code built on Ariel OS, which itself builds on Embassy, proved straightforward to port across hardware targets. VDL-Rust was successfully ported to an ST NUCLEO-F401RE with an IKS4A1 MEMS Expansion Board [24] with minimal changes, the majority of which concerned board-specific configuration details.

The equivalent VDL-C port required initializing a new project for the target board, copying portable application code, and defining pin mappings via preprocessor \#define directives. Where C-based projects require a separate project setup and manual code copying per target, Rust on Ariel OS consolidates everything within a single project — board-specific details are handled through \#[cfg] 
feature gates, and switching targets requires only a change to the build command.

Typically, this level of portability comes with significant drawbacks in binary size or execution speed. However, as we show in section VII, measurements against VDL-C on the same hardware indicate that these penalties are not substantial compared to traditional C runtimes used in the same context.

\subsection{Use of Async/Await}

The team implementing VDL-Rust was tasked to use Embassy and Ariel OS, which made using async/await the default choice.  While the first iterations were rapidly developed to working state, the tradeoffs and pitfalls of async Rust were not fully considered. This led to comparatively large code, while not making use of actual concurrency. Thus, tuning async/await required additional time and iterations to reach well performing code, as described in section VII.

In hindsight, a prior study of async Rust in more details would have enabled the Rust team to align the architecture with async Rust's constraints early on, avoiding some of the optimization effort required later in the project.

Even more radically: using regular synchronous code instead of async Rust was not considered, but would have been an option. 

\subsection{Avoiding Rust Binary Size Pitfalls}

{\bf Rust async/await Tradeoffs —} Async/await simplifies code, but each async fn compiles down to a state machine whose size grows with the number of local variables kept alive across await points — and this compounds with nested async calls, quietly inflating static RAM usage.

{\bf Enums Pitfalls —} Enums exacerbate the inflation we just described. The compiler allocates space equal to the largest variant, and when an enum is live across a suspension point, that full size is reserved for every await where it remains in scope.

{\bf Stack Frames —} To keep stack frames lean, large objects should be placed in .bss using StaticCell [27] rather than allocated as local variables. For large objects, init\_with() should be preferred over init(), as it constructs the value in-place within the StaticCell and avoids intermediate stack allocation.

{\bf Large struct Pitfalls —}  Passing a large struct by value into an initialization function may result in multiple copies on the stack before the value reaches its final location. Similarly, holding a large local variable and passing a \&mut reference to it keeps the backing storage on the stack for the variable's entire lifetime — the reference itself is cheap, but the stack allocation it points to is not.

\subsection{Avoiding Rust Execution Speed Pitfalls}

{\bf Rust async/await Latency —} Task-switch overhead with async/await is cycle-based and therefore clock-dependent. On the SensorTile.box Pro (160 MHz), a minimal switch — Timer::after\_micros(0).await — measures 1.7 µs (~272 cycles); accumulated across 2–3 switches reaches the observed 3.5–4 µs. The same 272 cycles consume ~3.2 µs at 84 MHz (STM32F401RE). This highlights that overhead budgets must be re-evaluated on each target, particularly at high ODR.

{\bf Peripheral Configurations —} Embassy's peripheral defaults prioritize cross-MCU compatibility over performance. Unlike STM32CubeMX, which generates target-specific initialization — maximum peripheral clocks, optimized I²C timings, tuned flash wait states — Embassy does not automatically replicate this envelope. Thus, for performance-sensitive ports of Embassy/Ariel OS, peripheral configurations should be cross-checked (e.g. against CubeMX output).

%% file: Text/8-1-related-work.tex
\section{Related Work}

Prior work such as [13] provides a survey of the embedded Rust landscape and related challenges. One such challenge is the size of Rust binaries which was so far larger than one might expect, as studied in [14]. On the other hand tangible workflow improvements of Rust firmware development compared to C/C++ firmware development was studied in [3] for Android devices. Concerning the microcontroller segment, a number of embedded software platforms entirely written in Rust have been recently developed to provide a basis for microcontroller firmware. For instance: TockOS [17] pioneered bare-metal, self-contained embedded Rust programming efforts. Another example is RTIC [16], a real-time interrupt-driven concurrency framework. Most recently, Embassy [8] and Ariel OS [2] provided embedded Rust software platforms leveraging async Rust and the rise of the crates ecosystem [18]. However, to the best of our knowledge, there was so far no comparative study of parallel Rust and C firmware development for microcontrollers, which is the focus here.

%% file: Text/9-conclusion.tex
\section{Conclusion}

For programming system software embedded on microcontrollers, both the actual performance and the technical readiness level of Rust have so far remained open questions. Answering these questions becomes urgent in a context where mean bugs keep showing up in large C/C++ code bases (predominantly used to date) while systems software must satisfy increasingly stringent requirements (for instance Cyber Resilience Act requirements). In this paper, we contributed some answers, by studying parallel efforts developing firmwares providing the same functionalities, respectively in Rust and in C, as required by an IoT industrial use case. Our analysis is based both on architectural considerations, on observations over the two teams’ iterative efforts, and on comparative performance measurements we carried out on common microcontroller hardware. The results of our study indicate that Rust is indeed a safe choice today. We also contribute lessons learned and overview the main pitfalls to be avoided by the upcoming generation of newbie Rust embedded developers on microcontrollers.

\section*{Code Availability}

The open source code of VDL-C/Rust is available online at \url{https://github.com/future-proof-iot/ST-Firmwares-VDL/}

\section*{Acknowledgements}
The authors thank Davide Aliprandi and Davide Sergi of the STAIoTCraft team, and the wider Ariel OS team. The work described in this paper was in parts financed by France 2030 via the PEPR Future Networks project FITNESS, and the PTCC project PQ-OTA.